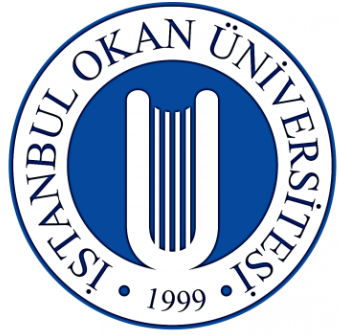

# HIGH INFLATION AND DEBT

## How Türkiye can overcome it?

**Abstract**

This report has been compiled by discussing inflation by knowing its history, by knowing its causes and considering its elimination.

Advisor: Prof. Dr. Fatma Çiğdem Çelik

Abdullah Mahmood
MBA BUSINESS ECONOMICS
222004041
abdmahmood@stu.okan.edu.tr

# Table of Contents

## Abstract:

The thing about inflation is that it ravages your income if you don't keep up with it and you don't know when it will stop. Emergency Measures were enacted in response to a rapid rise in this statistic and it requires higher interest rates which will raise borrowing costs significantly. In this report we will research on inflation in Turkiye. We will talk about its history and after that we will discuss the solution and how to control inflation and strengthen the economy. This research is done by collecting information from Government records, monetary records, World Bank and Organization for Economic Co-operation and Development (OECD).

## Introduction:

With a GDP of $720 billion, Turkey is the 19th largest economy in the world according to the World Bank report. Turkey is a member of OECD and G20. According to the World Bank, Turkey's commitments in fiscal year 2022 were 1591 million dollars.

Turkey's economy has experienced strong growth in recent years, with real GDP growing by an average of around 5% per year between 2002 and 2018. However, the economy has faced a number of challenges in recent years, including high inflation, a large current account deficit, and increased external debt. Inflation has been a persistent problem in Turkey, reaching a high of 25% in 2018. The current account deficit, which measures the difference between a country's imports and exports, has also been a concern, reaching a peak of nearly 7% of GDP in 2018.

Turkey has experienced strong economic growth in recent decades, with real GDP growth averaging around 5% per year in the 2000s. However, the economy has faced challenges in recent years, including high inflation, a large current account deficit, and a depreciating currency. The government has implemented a number of economic reforms in recent years, including measures to reduce the current account deficit and inflation, and to improve the business environment. These efforts have helped stabilize the economy and spur growth, but more work is needed to address structural challenges and further improve the business climate.

Overall, Turkey's economic outlook is mixed, with the potential for continued growth but also significant challenges and uncertainties to be addressed.

Economic indicators in Turkey have steadily improved over current-term Prime Minister Erdoğan's decade-long tenure, principally due to a boom on a series through considerably easing monetary policy, including lower benchmark rates, quantitative easing (QE) operations, asset purchases, and exchange rate guarantees across high-growth sectors such as construction and manufacturing. But new polices are not so strong, this year's 2022 inflation rate is 56.5%. And economists predicated 80% inflation rate for the upcoming year 2023.

Turkey is a diverse and dynamic economy that has undergone significant transformation in recent decades. Located at the crossroads of Europe and Asia, Turkey has a strategic position that has made it an important hub for trade and commerce. The country has a mixed economy that combines elements of both market and planned systems, and it has made significant progress in recent years in terms of economic development and liberalization.

It has a large and diversified industrial base, with leading sectors including textiles, automotive, chemicals, and construction. The country is also a major producer of agricultural products, including cereals, fruits, vegetables, and livestock. Tourism is another important sector of the economy, with the country's rich cultural heritage and diverse landscapes attracting millions of visitors each year.

It has a relatively young and educated population, which has helped to drive economic growth and development. The country has made significant investments in education and training, and it has a large and growing workforce that is well-suited to the demands of the modern economy.

The Turkish economy has faced a number of challenges in recent years, including high levels of inflation and public debt, as well as a reliance on external financing and a volatile currency. However, the government has implemented a number of structural reforms in recent years to address these issues, and the economy has shown signs of stabilization and recovery.

Overall, Turkey is a dynamic and diverse economy with significant potential for growth and development. It has a strategic location, a large and diverse industrial base, and a well-educated and youthful population, all of which make it an attractive destination for investment and trade.

The economy of any country is the guarantee of its growth or decline. If it takes ten years to build the best economy, then every year is necessary for a lifetime to maintain it. Years of hard work are lost because of just a few years of wrong policies. In this report I will take a detailed look at the Turkish economy. I will discuss the past, describe the present and what may happen in the future and mention the future plans of the Turkish government. In this report, there will be discussion on inflation, price review, private sector debt and solutions to all these problems. Turkey is one of the fastest growing economies in Europe, but also a contender with one of the highest rates of inflation. Turkey Economy focuses on Turkey's future growth potential and reports both positive and negative development. This can hinder their growth for years to come.

The discussion on Covid-19 is really necessary here. Because due to Covid-19 the inflation rates increases all over the World, Economy growth crashed and there is recession all over the World. Some countries made policies effective enough to get less shock from Covid-19. We will discuss the Turkiye Covid-19 policies here.

## Covid:

During the Covid period, the lira suffered a lot of depreciation against foreign currencies. The effect of which was that the investment in Turkey decreased considerably, but it would not be wrong to say that it was reduced to zero. But within the same year and the following year, the burden on the economy was reduced due to the policies made by the Turkish government regarding foreign investment. If we see, the positive trend in the economy was seen when the economy of quite strong countries was also vacillating. As in America and England, the economic condition during Covid was more down than in Turkey.

Covid-19, also known as the coronavirus disease, is a highly infectious and potentially deadly disease caused by the SARS-CoV-2 virus. It first emerged in Wuhan, China in late 2019 and has since spread to become a global pandemic, affecting nearly every country in the world.

The economic impact of Covid-19 has been severe, with many countries experiencing significant disruptions to their economies and labor markets. The effects have been felt most keenly in sectors such as tourism, hospitality, and retail, which have all seen a sharp decline in demand as a result of travel restrictions and lockdowns implemented to contain the spread of the virus.

The impact on the global economy has been widespread, with many countries experiencing significant declines in GDP (gross domestic product), high unemployment rates, and increased government debt. Many businesses have been forced to close or scale back operations, leading to widespread job losses and economic hardship for individuals and families.

Governments around the world have implemented a range of measures to support their economies and mitigate the impact of the pandemic, including fiscal measures such as stimulus packages and emergency funding, and monetary measures such as interest rate cuts and asset purchases.

Despite these efforts, the economic effects of Covid-19 are likely to be felt for some time, and it is uncertain what the long-term impact on the global economy will be. It is important for governments and businesses to continue to work together to find ways to support economic recovery and ensure a sustainable future for all.

Covid was something that shook the whole world. Turkey is also among the countries that have been badly affected by Covid. Due to Covid, the economy was badly hit. The prices of things went up very quickly. People's purchasing power has become negligible in 2021. With the passage of time as the country came out of the lockdowns and ill effects of Covid and slowly everything became normal. But now that 2022 is coming to an end, there are still many problems caused by Covid. In 2020, when the Covid had fully effected in Turkey, the level of inflation that occurred was not possible to recover until now. Rather, it is increasing day by day. Yes, it is a welcome thing that the future plan and the economy will improve significantly in the coming year 2023. The Turkish lira will be supported. But it is important to note that as long as the dollar continues to support the economy, the Turkish lira will continue to depreciate.

The banking crisis of 2001 was a great lesson. From which the Turkish government learned a lot. Then the Turkish lira devalued by 6% and GDP fell below 5%. Then Turkey had an agreement with the IMF. Due to this agreement, all the working departments of Turkey have to work transparently. Due to which the level of corruption has reduced significantly. Within five years after the 2001 crisis, Turkey recorded a growth rate of seven percent. Then the government insisted on setting up the industry. As new factories were installed, Turkey moved up the list of industrialized countries.

## Turkiye Covid Policies:

As can be seen in the figure below, the dollar had to be used and foreign reserves dwindled. CPI and exchange rate remained on the decline. Due to which people had to face a lot of difficulties. Injecting the dollar into the market to keep the lira artificially strong proved to be a difficult and time-consuming decision. The economy was strong for a while, but for the long term or two years, this plan was the weakest. That's why prices started to rise and foreign reserves also got low. Turkey faced a new challenge. And that challenge was rising inflation and loans to the private sector. If we consider Turkey's 2030 plan, the focus is on overcoming this very challenge.

BRSA showed 100 to 95% of assets as of August 2020. The same ratio has been more difficult since January 2021. NPL, i.e. non-performing loans, faced almost 70% of the private sector during Covid. As a result (as can be seen in the figure below) the NPL was capped at five percent. This provided enough headroom to protect the economy from being disrupted during the lockdown. Private debt was 58.72 percent of GDP in September 2020. Which is 41.45 percent in September 2022 this year. It happened in a low record, but it has been a year and a half since the lockdown and the economic crisis. In this sense, this debt is very high. Below is the graph showing private debt percentage of Gross domestic product. We clearly can see the changes during each year. This graph is made while BRSA statistics under observation.

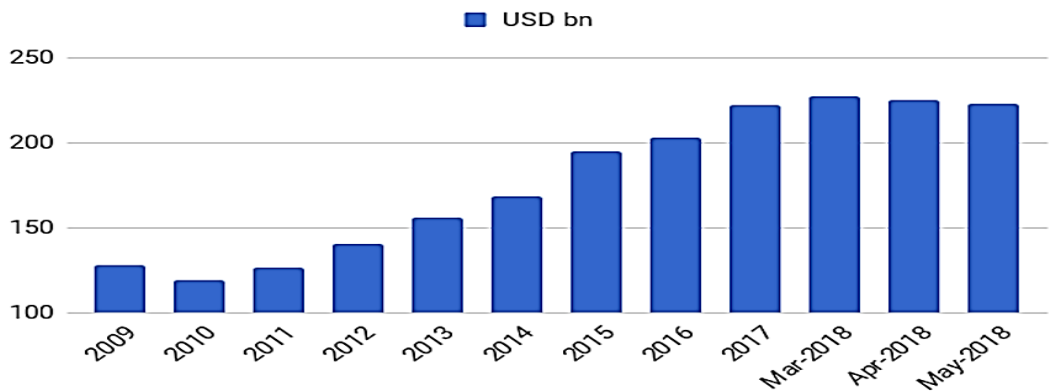

## In 2020 to 2022:

In 2009 the foreign private debt was $278,829,978,028. Which increased nearly every year. In 2015 this was $399,948,917,893. And in 2020 it was $435,889,447,921. In 2022 which is going to and end after this month Dec 2022 the foreign private debt is 444.4 billion USD in Jun 2022, compared with 450.5 billion USD in the previous quarter.

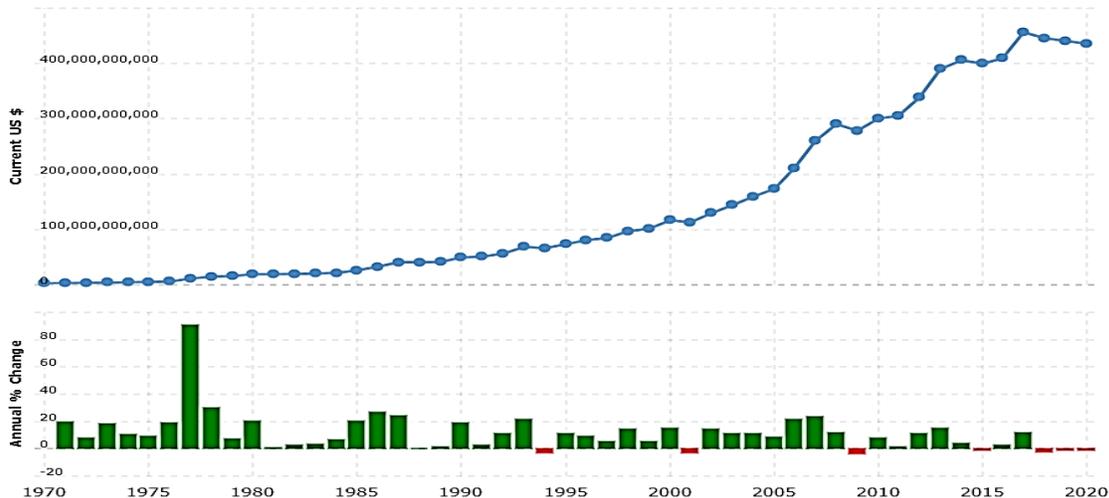

## Problem is:

Turkey's economy was affected by the financial crisis that started with US housing problems. As this led to increased unemployment and bankruptcy rates, Turkish household expenditure decreased and had an immense impact on other sectors of their export market like construction and textile industries. As global trade fell, inflationary rate rose sharply in 2020 which paved the way for new high interest rates to be applied in 2021-22 which stalled inflation rise as well as slowed down growth in retail sector. Turkey has also faced connectivity-related issues because it has faced international isolation such as withdrawal from a free trade agreement with Russia and Europe's refusal to participate in major nuclear power plants being constructed by Turkey. Inflation and Inflation uncertainty.

The Turkish lira has also faced significant depreciation against major currencies in recent years, which has contributed to the country's high inflation and current account deficit. The government has implemented a number of measures to address these challenges, including tightening monetary policy, implementing structural reforms, and improving the business environment.

Inflation creates distrust in the society. Educated and skilled people start migrating to other countries. Where inflation is low, business or job opportunities are high. This is why inflation and Inflation uncertainty create distortions in society. The economy of the country is slowly consumed. Turkey has a long and volatile history of inflation that has increased over time. If we add up the statistics, the period of inflation from 1974 to 2004 appears to be the fastest. From 2004 to 2007, inflation in Turkey was well under control.

The graph below has been dropped to show clearly the monetary policy of Turkey. Structural changes can be clearly noted. This period from 1975 to 2017 is of special importance. If we study from 1975 to 1980, the main cause of inflation was the increase in oil prices. Inflation from 1980 to 2000 was caused by balance of payments crisis. Since 1980, when Turkey signed up to the IMF, inflation was raised to pay back the loan that Turkey received under the Joint Fund. This burden fell on the people.

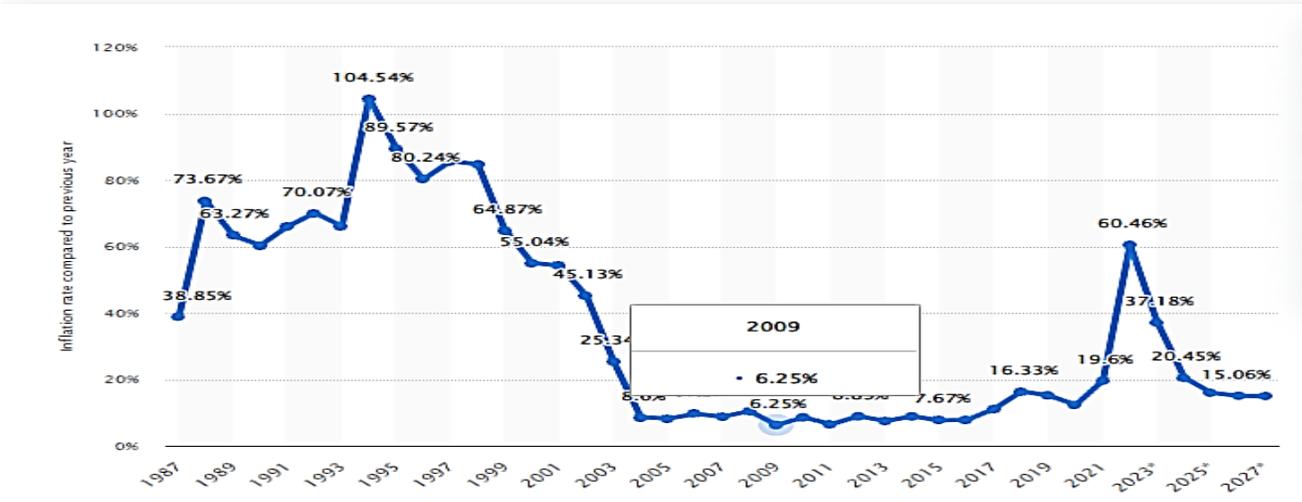

Which was inevitable. As a result, inflation increased, but after the loan, because there were funds in foreign reserves, the increase was recorded at a higher rate. The inflation rate from 1971 to 1980 was 33.6%.

After capital account liberalization, Turkey once again faced financial crisis. This period was from 1994 to 2001 when the inflation rate reached 77.2%. Turkey undertook economic stabilization programs in the 90's but they were not very successful. To deal with these crises, Turkey once again went to the IMF. From 2002 to 2007, the IMF program was implemented and inflation dropped to just 7%. Inflation is of great importance under macroeconomic conditions in Turkey. This is the reason why this report has been compiled by discussing this inflation by knowing its history, by knowing its causes and considering its elimination.

## The Unemployment:

Total population of Turkiye is 86,559,586 as of Thursday, December 15, 2022. In this year December 2022, the employment rate in Turkey is at 10.2%. Which was 10.1 in October 2022. The number of these employed people has increased by 57 thousand to more than 35.35 million. While the number of employed people increased from 229 thousand to 31.200 million.

When unemployment increases in a country, the economy is badly affected. Inflation and unemployment are closely related. Skilled and educated people move to other countries when they do not find work. This makes the country hollow. Also, other countries do not invest in your country because they feel the risk of not getting profit. There is uncertainty in the country. The graph below gives a good idea of how big the implementation is in Turkey. Which is also seeing some improvement in the past. But if the positive ratio is seen, the business has increased. A relaxation of the loan system and bank conditions brought in new investors and created jobs. This has also led to a significant reduction in the unemployment rate.

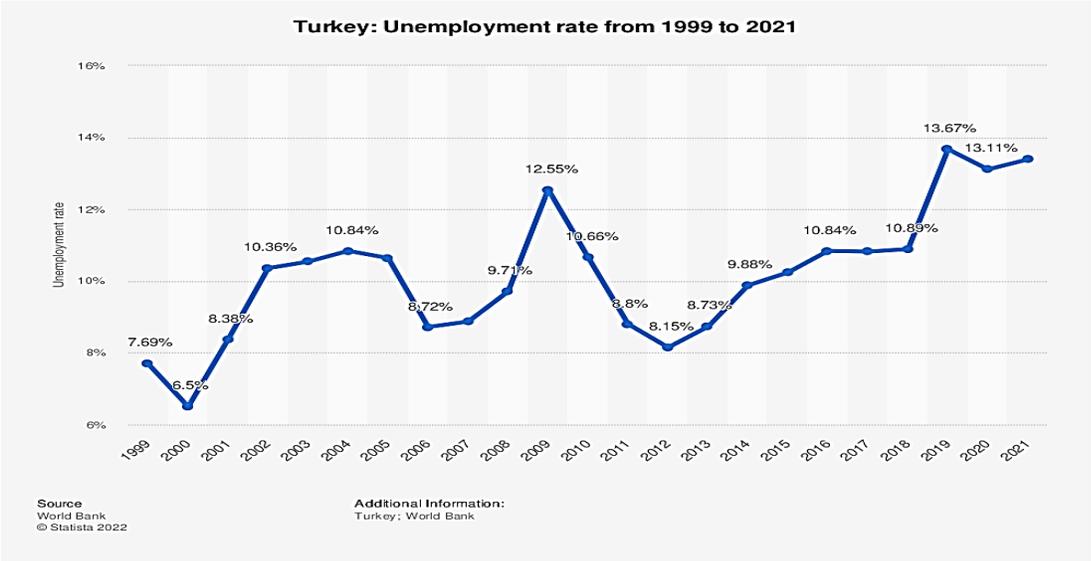

## Solution for Unemployment:

Unemployment refers to the condition of people who are able to work but are unable to find jobs. It is a measure of the health of the labor market and can have significant economic and social consequences.

The main reasons for unemployment are lack of skills, giving temporary jobs, working on probation, not being more active in the law, not having equality in gender, less practical in universities and educational institutions and more work on theory. And the government is not paying enough attention to this issue.

If we look at the graph carefully, we will see that the ratio was 9% in 2001 but increased to 16% in 2002. In 2010, this ratio had reached 35.5%. In 2022, it decreased significantly. But still inflation is increasing very fast due to which small and medium size businesses are closing down. And there are chances of considerable increase in unemployment by next year.

The solution to unemployment is focusing on new opportunities for people. Built new industries. Produce common to advanced products in own country. Turkiye economy from 2001 to 2010 was same to promote the industries in the country. This is the solution to raise the jobs in the country. Producing things, basic commodities or daily life commodities will not just increase the skillful people in the country, decrease unemployment ratio but also decrease the imports. This will strengthen the economy overall. Apart from promoting industrial policies. Education organizations should promote skills than theories in their education system.

Tax relaxation to foreign investors also a solution to increase employment in the countries. When foreign countries invests, industries promoted in the country. New organization produces a lot of jobs. This is really valuable practice which is done by India from 2015 to 2020. Technology giants such as Apple, Google, Microsoft, HP, Dell and Intel invested in India. Because Indian government gave tax relaxation to these multi-billion companies. At date Indian foreign reserves are more than 500Billion dollars and within five years of this policy more than 12 million people employed.

The law should be more active. Job security should be make sure in private sector across country. If social security system strengthened in the country, temporary and probation jobs will be discouraged. The uncertainty in jobs creates really upset situation in young generation. There are several ways in which governments can try to address unemployment. Some of the most common approaches include:

**Fiscal policy:** Governments can use fiscal policy, such as adjusting tax rates or government spending, to try to stimulate economic growth and create new jobs.

**Monetary policy:** Central banks can use monetary policy, such as adjusting interest rates, to try to influence economic activity and employment.

**Job training and education programs:** Governments can invest in programs that provide job training and education to help workers acquire the skills needed to find employment.

**Unemployment insurance:** Governments can provide unemployment insurance to provide temporary financial assistance to workers who have lost their jobs.

**Public works projects:** Governments can invest in public works projects, such as infrastructure projects, to create jobs and stimulate economic growth.

It's worth noting that addressing unemployment can be a complex and challenging task, and different approaches may be more or less effective depending on the specific circumstances of a given economy.

## Inflation:

Inflation is an increase in the general price level of goods and services in an economy over a period of time. When the general price level rises, each unit of currency buys fewer goods and services; consequently, inflation reflects a reduction in the purchasing power of money – a loss of real value in the medium of exchange and unit of account within an economy. A chief measure of price inflation is the inflation rate, the annualized percentage change in a general price index (normally the consumer price index) over time.

Inflation in Turkey reached a peak of over 100% in the late 1970s and early 1980s, but it has generally trended downwards since then. However, inflation in Turkey has remained relatively high compared to other countries, and it has been prone to spikes in recent years. In 2021, for example, inflation in Turkey reached almost 20%.

Inflation in Turkey has been a persistent problem for many years. The country has a history of high and volatile inflation, which has had a negative impact on the economy and the living standards of Turkish citizens. Inflation in Turkey has been driven by a number of factors, including structural economic problems, political instability, and external economic shocks. Some of the key factors that have contributed to high and volatile inflation in Turkey include:

**Fiscal deficits and public debt:** High levels of government spending and persistent budget deficits have contributed to inflation in Turkey.

**Exchange rate instability:** Turkey has a floating exchange rate, which means that the value of the Turkish lira is determined by market forces. Exchange rate instability has contributed to inflation in Turkey by making imported goods more expensive.

**Political instability:** Political instability has often led to economic instability in Turkey, which has in turn contributed to inflation.

**Structural economic problems:** Turkey has struggled with structural economic problems, including low savings and investment rates, which have contributed to low productivity and high inflation.

**Solution for the inflation:**

There are several ways to address the problem of inflation:

**Monetary policy:** Central banks, such as the Federal Reserve in the United States, can use a variety of tools to influence the supply and demand of money and credit in the economy, which can help to control inflation. For example, the central bank can raise interest rates to reduce the supply of money and credit, which can help to reduce inflation.

**Fiscal policy:** Governments can use fiscal policy, such as changes in taxation and government spending, to influence the level of demand in the economy and help to control inflation. For example, the government can increase taxes or reduce spending to reduce demand and help to reduce inflation.

**Price controls:** Governments can also use price controls, such as setting maximum prices for certain goods and services, to try to curb inflation. However, price controls can have unintended consequences and may not be effective in the long run.

**Exchange rate policy:** A country's exchange rate, or the value of its currency in relation to other currencies, can also affect inflation. If a country's exchange rate is overvalued, its exports may become less competitive, leading to slower economic growth and lower inflation. Conversely, if a country's exchange rate is undervalued, its exports may become more competitive, leading to faster economic growth and higher inflation.

**Supply-side policies:** Governments can also implement policies that aim to increase the supply of goods and services in the economy, such as by investing in infrastructure and education or by reducing regulations that may be hindering economic growth. These policies can help to increase productivity and reduce inflationary pressures.

**Private Debt:**

Private debt refers to the debt incurred by private sector entities, such as households, businesses, and non-governmental organizations. Private debt can take many forms, including mortgages, credit card debt, and business loans. Private debt can have both positive and negative impacts on the economy. On the one hand, it can help individuals and businesses to finance investments and consumption, which can stimulate economic growth. On the other hand, excessive private debt can be a source of financial instability, as it can lead to defaults and bankruptcies if borrowers are unable to meet their debt obligations. This can have negative impacts on the economy, as it can reduce the availability of credit and discourage investment.

In Turkey, private debt has been a significant factor in the country's economic growth in recent years. For example, the Turkish government has encouraged the expansion of mortgage lending, which has helped to boost the housing market and spur economic growth. However, the growth in private debt has also raised concerns about the sustainability of the country's economic growth and the potential for a debt crisis.

The high level of private debt has been a concern for policymakers, as it has contributed to the country's high current account deficit and made the economy more vulnerable to external shocks. The government has implemented a number of measures to address this issue, including tightening monetary policy, improving the business environment, and implementing structural reforms to encourage economic growth.

In 2018, Turkey experienced a financial crisis, partly due to high levels of private debt. The crisis was triggered by a combination of factors, including a large current account deficit, high levels of external debt, and a depreciation of the Turkish lira. The crisis led to a sharp contraction in the economy, with real GDP falling by 2.4% in 2018 and by 1.8% in 2019.

To address the crisis, the Turkish government implemented a number of measures, including tightening monetary policy, implementing structural reforms, and improving the business environment. These efforts have helped to stabilize the economy, and real GDP has been growing again since 2020. However, the high levels of private debt and the ongoing challenges faced by the economy remain a concern for policymakers and investors.

## Trades and Foreign reserves:

Foreign reserves are a country's foreign currency deposits and bonds held by central banks and monetary authorities. These reserves can be used to finance international trade, as well as to stabilize their own currency.

Foreign reserves are important for a number of reasons. They can help a country to:

**Maintain a stable exchange rate:** A country with a large amount of foreign reserves can use them to stabilize the value of its currency by buying or selling its own currency on the foreign exchange market.

**Meet external financial obligations:** Foreign reserves can be used to pay for imported goods and services, as well as to service external debt.

**Protect against external shocks:** Foreign reserves can act as a buffer against external economic shocks, such as a sudden decline in the value of the country's own currency or a drop in the price of its exports.

**Promote international trade:** Foreign reserves can help to facilitate international trade by providing a means of exchange for goods and services.

In summary, foreign reserves are an important component of a country's economic stability and play a key role in maintaining a healthy balance of payments. They can also contribute to the overall stability of the international monetary system.

Turkey is a country with a diverse economy that is heavily reliant on foreign trade. The country's main exports include textiles, clothing, automotive parts, iron and steel products, and agricultural products such as fruits and vegetables. Turkey's main trading partners include the European Union, Russia, China, and the United States.

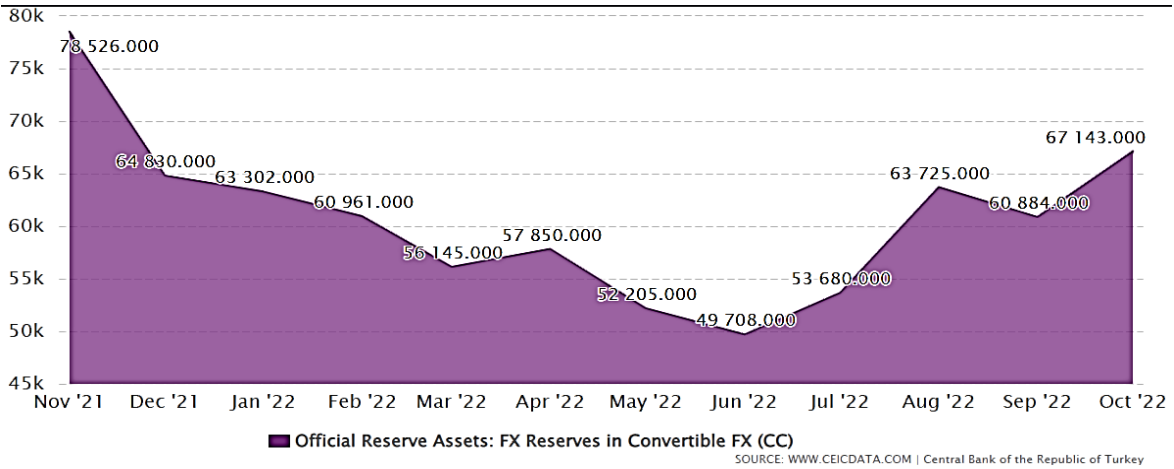

Turkey's foreign reserves consist of a variety of assets, including gold, foreign currency, and financial instruments such as bonds and securities. The country's foreign reserves are managed by the Central Bank of the Republic of Turkey, which uses them to help stabilize the exchange rate and maintain financial stability.

The level of Turkey's foreign reserves has fluctuated over time. In recent years, the country has faced challenges due to economic challenges such as high inflation, a large current account deficit, and a depreciation of the Turkish lira. These challenges have led to a decline in the level of Turkey's foreign reserves, which have fallen from a peak of over $200 billion in 2014 to around $80 billion in 2021.

Despite these challenges, Turkey's economy remains strong, and the country has continued to engage in international trade and investment. Turkey is a member of the World Trade Organization and has free trade agreements with a number of countries, including the European Union, which has helped to increase its competitiveness in international markets.

## Foreign Trades are important?

Foreign trade, or international trade, refers to the exchange of goods and services between countries. It is a key aspect of the global economy and can bring many benefits to participating countries.

There are several ways in which foreign trade can be beneficial for a country's economy:

**Increased specialization:** Foreign trade allows countries to specialize in producing goods and services in which they have a comparative advantage, meaning they can produce these goods and services at a lower cost or higher quality than other countries. Specialization can lead to increased efficiency and productivity, which can benefit the economy by increasing the overall supply of goods and services available to consumers.

**Access to new markets:** Foreign trade can also provide access to new markets for a country's goods and services. This can be particularly beneficial for small and medium-sized enterprises that may not have the resources to tap into domestic markets. By exporting their goods and services to other countries, these businesses can increase their revenue and contribute to the growth of the economy.

**Increased competition:** Foreign trade can also increase competition within a country by exposing domestic producers to competition from foreign producers. This can encourage domestic firms to become more efficient and innovative in order to compete with foreign firms, leading to increased productivity and economic growth.

**Improved living standards:** Foreign trade can also contribute to improved living standards in a country by providing access to a wider variety of goods and services at lower prices. This can lead to increased purchasing power for consumers and a higher standard of living.

It's important to note that while foreign trade can bring many benefits to a country's economy, it can also have some negative effects. For example, foreign competition can lead to job losses in certain sectors, and trade policies can sometimes have unintended consequences for certain groups of people or industries. It's important for governments to carefully consider the potential impacts of foreign trade on all sectors of the economy and to take steps to mitigate any negative effects.

## Foreign Reserves are important?

Foreign reserves are important for a country for a number of reasons. Here are some of the main reasons why foreign reserves are important:

**Foreign reserves provide a buffer against economic shocks:** Foreign reserves can be used to stabilize a country's exchange rate and provide a buffer against economic shocks, such as a sudden drop in the value of the country's currency or a sudden decrease in the demand for its exports. By having a large pool of foreign reserves, a country can use these reserves to stabilize its exchange rate and maintain the value of its currency.

**Foreign reserves can be used to finance imports:** Foreign reserves can be used to finance a country's imports, which are essential for the country's economic growth and development. By having a sufficient amount of foreign reserves, a country can ensure that it has the resources it needs to import the goods and services it needs to meet the demands of its domestic market.

**Foreign reserves can improve a country's creditworthiness:** Foreign reserves can also improve a country's creditworthiness, as they can be used to back the country's financial obligations. This can make it easier for the country to borrow money from international financial institutions or other countries, which can be important for financing infrastructure projects or other development initiatives.

**Foreign reserves can increase investor confidence:** Finally, foreign reserves can also increase investor confidence in a country's economic stability and prospects. By having a large pool of foreign reserves, a country can demonstrate to investors that it has the financial resources to meet its financial obligations and weather economic challenges. This can make it more attractive for investors to invest in the country, which can help to stimulate economic growth and development.

## Turkish Lira vs American Dollar:

The value of a currency, such as the Turkish lira or the American dollar, is determined by a variety of factors, including economic conditions, political stability, and the level of demand for the currency in international financial markets.

In general, the value of a currency is determined by the supply and demand for that currency in the foreign exchange market. When there is a high demand for a currency, its value tends to increase relative to other currencies. Conversely, when there is a low demand for a currency, its value tends to decrease.

There are several factors that can affect the demand for the Turkish lira or the American dollar. For example, economic conditions in a country can influence the demand for its currency. If a country has a strong and growing economy, this may increase the demand for its currency, as investors and businesses are more likely to want to hold and use the currency. On the other hand, if a country has a weak or declining economy, this may decrease the demand for its currency, as investors and businesses may be less likely to want to hold or use it.

Political stability can also affect the demand for a currency. If a country has a stable political environment, this may increase the demand for its currency, as investors and businesses are more likely to feel comfortable holding and using the currency. Conversely, if a country has a volatile or uncertain political environment, this may decrease the demand for its currency, as investors and businesses may be less likely to feel confident holding or using it.

In addition to economic and political factors, the level of demand for a currency in international financial markets can also affect its value. For example, if there is a high level of demand for the Turkish lira or the American dollar from foreign investors or businesses, this may increase the value of the currency relative to other currencies. On the other hand, if there is a low level of demand for the Turkish lira or the American dollar, this may decrease its value.

It's also worth noting that the value of a currency can be influenced by the actions of central banks, such as the US Federal Reserve or the Central Bank of the Republic of Turkey. For example, if a central bank raises interest rates, this can increase the demand for its currency, as investors may be more likely to hold the currency in order to earn higher returns on their investments. Conversely, if a central bank lowers interest rates, this can decrease the demand for its currency, as investors may be less inclined to hold the currency in order to earn lower returns on their investments.

## Conclusion:

There are many factors that can influence the economy of a Turkiye, and the specific steps that can be taken to improve it will depend on the unique circumstances I mentioned in the report. However, here are some general strategies that can be considered:

**Promote economic growth:** This can be done through investments in infrastructure, education, and research and development. Encouraging businesses to start and expand can also help drive economic growth.

**Foster a supportive business environment:** This can include reducing red tape and bureaucracy, simplifying tax codes, and protecting property rights.

**Foster international trade:** Encouraging exports and imports can help stimulate economic growth by providing access to a larger market for goods and services.

**Promote fiscal responsibility:** This can involve reducing government debt, maintaining low inflation, and ensuring that government spending is focused on productive investments.

**Invest in human capital:** Providing access to education and job training can help increase productivity and competitiveness.

**Foster innovation:** Encouraging entrepreneurship and supporting research and development can help drive productivity and economic growth.

**Address income inequality:** Reducing income inequality can help ensure that the benefits of economic growth are more widely shared and can help stimulate demand for goods and services.

**Foster a stable financial system:** A stable financial system is important for providing the credit and financing needed for businesses to invest and grow.

**Foster a stable political environment:** Political stability can help create a sense of predictability and confidence, which can encourage investment and economic growth.

**Promote sustainable development:** Ensuring that economic growth is sustainable and does not harm the environment can help ensure long-term prosperity.

It's important to note that improving the economy of a country is a complex process and requires a multifaceted approach. The specific steps that a country should take will depend on its unique circumstances and challenges. All major challenges are discussed in this report.

[i] All information in this report is cross-checked from multiple resources during the research.